\documentclass[reprint,superscriptaddress,aps,pra]{revtex4-1}
\usepackage{setspace}
\usepackage{amsmath}
\usepackage{amsfonts}
\usepackage{amssymb}
\usepackage{graphicx}
\usepackage{braket}
\usepackage{hyperref}
\usepackage[english]{babel}
\usepackage{color}
\usepackage[textsize=scriptsize]{todonotes}
\usepackage{siunitx}
\newcommand{\Om}{\Omega_{\mathrm{m}}}
\newcommand{\Gm}{\Gamma_{\mathrm{m}}}
\newcommand{\Sm}{\Sigma_{\mathrm{m}}}

\newcommand{\oo}{\omega}

\newcommand{\opr}{\omega_\mathrm{p}}
\newcommand{\octrl}{\omega_\mathrm{L}}

\newcommand{\Soo}{\Sigma_{\mathrm{o}_1}}
\newcommand{\Sot}{\Sigma_{\mathrm{o}_2}}
\newcommand{\kk}{\kappa}

\newcommand{\Dbaronetwo}{\bar{\Delta}_{1,2}}
\newcommand{\Db}{\bar{\Delta}}

\newcommand{\C}{\mathcal{C}}

\newcommand{\T}{^{\text{T}}}

\newcommand{\as}{^\ast}
\newcommand{\dg}{^\dagger}
\newcommand{\im}{\mathrm{i}}

\begin{document}
\title{Optical circulation in a multimode optomechanical resonator}
\author{Freek Ruesink}
\thanks{These authors contributed equally to this work.}
\affiliation{Center for Nanophotonics, AMOLF, Science Park 104, 1098 XG Amsterdam, The Netherlands}
\author{John P. Mathew}
\thanks{These authors contributed equally to this work.}
\affiliation{Center for Nanophotonics, AMOLF, Science Park 104, 1098 XG Amsterdam, The Netherlands}
\author{Mohammad-Ali Miri}
\affiliation{Department of Electrical and Computer Engineering, The University of Texas at Austin, Austin, TX 78712, USA}
\author{Andrea Al\`{u}}
\affiliation{Department of Electrical and Computer Engineering, The University of Texas at Austin, Austin, TX 78712, USA}
\author{Ewold Verhagen}
\email{verhagen@amolf.nl}
\affiliation{Center for Nanophotonics, AMOLF, Science Park 104, 1098 XG Amsterdam, The Netherlands}
\date{\today}

\renewcommand{\figurename}{Fig.}
\renewcommand{\refname}{References and Notes}

%SciAdv - Abstract not to exceed 250 words and ideally closer to 200. Do not include citations.
\begin{abstract}
Optical circulators are important components of modern day communication technology. With their ability to route photons directionally, these nonreciprocal elements provide useful functionality in photonic circuits and offer prospects for fundamental research on information processing.
Developing highly efficient optical circulators thus presents an important challenge, in particular to realize compact reconfigurable implementations that do not rely on a magnetic field bias to break reciprocity.
We demonstrate optical circulation based on radiation pressure interactions in an on-chip multimode optomechanical system. 
We show that mechanically-mediated optical mode conversion in a silica microtoroid provides a synthetic gauge bias for light, which enables a 4-port circulator by exploiting tailored interference between appropriate light paths. We identify two sideband conditions under which ideal circulation is approached. This allows to experimentally demonstrate $\sim$10 dB isolation and $<$~3 dB insertion loss in all relevant channels.
We show the possibility of actively controlling the bandwidth, isolation ratio, noise performance and circulation direction, enabling ideal opportunities for reconfigurable integrated nanophotonic circuits.
\end{abstract}
 
\maketitle

\section*{Introduction}
\noindent{}Optical circulators route photons in a unidirectional fashion among different ports, with diverse applications in advanced communication systems, including dense wavelength division multiplexing and bi-directional sensors and amplifiers. Their operation offers opportunities for routing quantum information~\cite{Sliwa2015,Scheucher2016} and to realize photonic states whose propagation in a lattice is topologically protected~\cite{Lu2014,Roushan2017}. Traditionally, nonreciprocal elements such as circulators and isolators have relied on applied magnetic bias fields to break time-reversal symmetry and Lorentz reciprocity. While significant progress has been made towards introducing magneto-optic materials in photonic circuits~\cite{Shoji2014}, realizing low-loss, linear, efficient, and compact photonic circulators on a chip remains an outstanding challenge. 
In recent years, coupled-mode systems that create an effective magnetic field using parametric modulations have been recognized as powerful alternatives~\cite{Fang2012,Poulton2012,Estep2014,Sliwa2015,Kim2015,Kerckhoff2015,Ranzani2015,Metelmann2015}.
Optomechanical systems, where multiple optical and mechanical modes are coupled through radiation pressure~\cite{Aspelmeyer2014,Hill2012,Andrews2014}, provide an effective platform to realize such modulation~\cite{Hafezi2012,Xu2015,Miri2017}.
Nonreciprocal transmission and optical isolation were demonstrated in several optomechanical implementations, including ring resonators, photonic crystal nanobeams, and superconducting circuits~\cite{Shen2016,Ruesink2016,Fang2017,Peterson2017,Bernier2016}.
Recently, nonreciprocal circulation for microwave signals was reported in a superconducting device exhibiting directional frequency conversion~\cite{Barzanjeh2017}. 
Here, by leveraging a synthetic gauge field created in a multimode optomechanical system, we realize a compact and highly reconfigurable on-chip circulator in the photonic (telecommunication) domain. 
The working principle, relying on tailored interfering paths, allows for operation at equal-frequency input and output fields and could be applied in a wide variety of platforms.
We reveal the importance of control fields to regulate the nonreciprocal response and identify two distinct regimes, with and without employing optomechanical gain, where this response can approach ideal circulation. 

%\colev{We explain that interference of nonreciprocal paths allows operation at equal-frequency input and output fields, and establish general working principles that could be applied in a wide variety of platforms. 
%We reveal the importance of control fields to regulate the nonreciprocal response and identify two distinct regimes, with and without employing optomechanical gain, where this response can approach ideal circulation. 
%For both, we explore isolation, loss, bandwidth, and noise performance.}

% \colev{
% The working principle, relying on interference between a direct and nonreciprocal path, allows for operation at equal-frequency input and output fields and could be applied in a wide variety of platforms. 
% We discuss the device properties, including its noise performance, which are shown to strongly depend on the applied control field regulating the nonreciprocal response.
% }

\begin{figure}
\centering
\includegraphics[width=88mm]{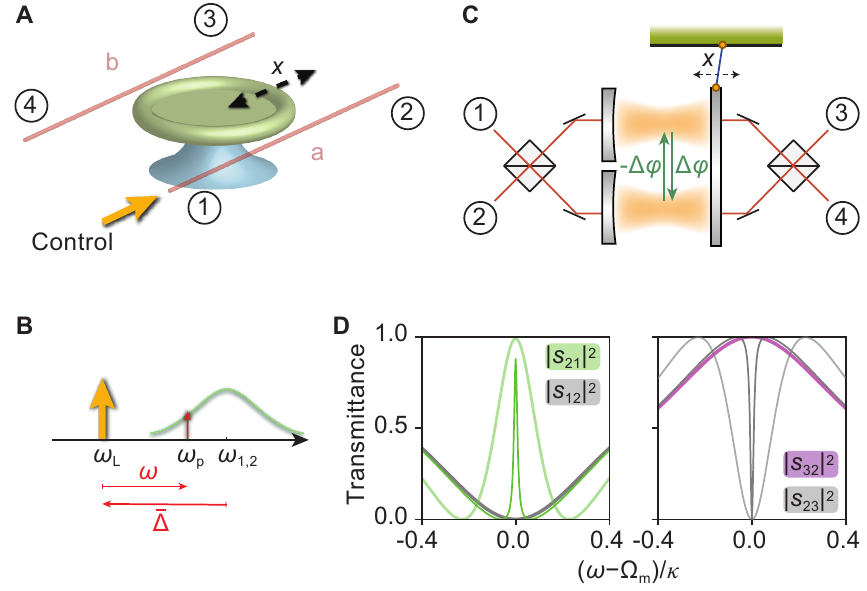}
\caption{\textbf{Optical nonreciprocity in a multimode optomechanical system.} 
(\textbf{A}) An optomechanical microtoroid coupled to optical fibres a (ports 1,2) and b (ports 3,4).
Even and odd optical whispering gallery modes interact with a mechanical radial breathing mode through radiation pressure.
A control beam, entering in port 1, can induce nonreciprocal transmission in the direct ($1\leftrightarrow2$, $3\leftrightarrow4$) and add-drop ($2\leftrightarrow3$, $4\leftrightarrow1$) channels. 
(\textbf{B}) Control ($\octrl$) and probe ($\opr$) frequencies with respect to the cavity frequencies $\oo_{1,2}$.
(\textbf{C}) An analogous Fabry-P\'{e}rot model for the three-mode system is shown. The optical mode conversion mediated by the mechanical mode imprints a nonreciprocal phase $\Delta\phi$ on forward and reverse intercavity mode transfers. 
(\textbf{D}) Theoretically predicted probe transmittances for the direct (left panel) and add-drop (right panel) channels for $\Db=-\Om$, negligible intrinsic losses and two different cooperativities \mbox{($\C=15$ and 200)}.
The nonreciprocal response is inferred from $s_{ij}\neq s_{ji}$. Its strength and bandwidth are increased through higher cooperativity (lighter shade).
}
\label{fig:schematic}
\end{figure}

\section*{Results}
\noindent{}Our experimental system consists of a high-Q microtoroid resonator~\cite{Armani2003} that is simultaneously coupled to two tapered optical fibres, forming four ports through which light can enter and exit (Fig.~\ref{fig:schematic}A). 
The microtoroid supports nearly degenerate odd and even optical modes --- superpositions of clockwise and counterclockwise propagating waves --- coupled through a mechanical breathing mode, to form a three-mode optomechanical system. 
A control beam incident through port 1 populates both modes, labelled by $i=\{1,2\}$, with intracavity control fields $\alpha_i$ that exhibit $\pi/2$ phase difference~\cite{Yu2009}.
%Importantly, this results in  biased and enhanced optomechanical coupling rates $g_i=g_0\alpha_i$, where $g_0$ is the vacuum optomechanical coupling rate~\cite{Aspelmeyer2014}\todo{TSDU}.
When detuned from the cavity, these control fields induce linear couplings between the mechanical resonator and both cavity modes at rates $g_i=g_0\alpha_i$, where $g_0$ is the vacuum optomechanical coupling rate~\cite{Aspelmeyer2014}.
For red-detuned control (Fig.~\ref{fig:schematic}B), photon transfer from mode 1 to mode 2 via the mechanical resonator then takes place at rate $\mu_\mathrm{m}=2g_1^\ast g_2/\Gm$, with $\Gm$ the mechanical linewidth. 
In contrast, transfer from mode 2$\rightarrow$1 occurs at rate $\mu_\mathrm{m}^\ast$, i.e., with opposite phase. 
The control field thus biases the mode-conversion process with maximally nonreciprocal phase $\Delta\phi\equiv\arg(\mu_\mathrm{m})=\pi/2$~\cite{Ruesink2016,Miri2017}. 
This phase is reminiscent of a synthetic d.c. magnetic flux that breaks time-reversal symmetry~\cite{Fang2012a} and can, under appropriate conditions, serve to create an optical circulator. 

Our system is fully equivalent to the general Fabry-P\'{e}rot model sketched in Fig.~\ref{fig:schematic}C, where the beamsplitters %that distribute light from the ports to the two cavities 
impart a $\pi/2$ phase shift on reflections from either side.
From each port the two cavities are then excited with the same $\pi/2$ phase difference as the even and odd modes in the ring resonator.
In absence of optomechanical coupling this realizes a reciprocal `add-drop' filter: off-resonant light propagates from port 1(3) to port 2(4) and vice versa, while resonant light is routed from 1(2) to 4(3).
However, in the presence of a control beam at a frequency $\octrl$ detuned from the cavity frequencies $\oo_{1,2}$ by $\Db=-\Om$, with $\Om$ the mechanical resonance frequency, a weaker probe beam at frequency $\opr=\octrl+\oo$ (Fig.~\ref{fig:schematic}B) experiences an optomechanically induced transparency (OMIT~\cite{Weis2010}) window when $\oo\approx\Om$: the probe is routed from port 1 to 2 rather than `dropped' in port 4.
This behaviour is quantified in Fig.~\ref{fig:schematic}D, which shows scattering matrix elements $s_{ij}$, signalling transmission from port $j$ to $i$, for two control powers. These are obtained from a rigorous coupled-mode model describing the optomechanical three-mode system (Materials and Methods).
The response is clearly nonreciprocal: the OMIT window for $s_{21}$ (green curves, left panel) does not occur for probe light incident from port 2 (grey), which remains dropped in port 3 (purple, right panel). Likewise, light from port 3 \emph{is} suppressed (grey) and routed to port 4, etc.
The optomechanical interactions in this multimode system thus allow nonreciprocal add-drop functionality, which can also be related to momentum-matching of intracavity control and probe fields~\cite{Hafezi2012}.
For proper bias conditions this yields a circulator, i.e., a control beam incident from port 1 or 3 allows probe transmission  1$\rightarrow$2, 2$\rightarrow$3, 3$\rightarrow$4, 4$\rightarrow$1, whereas a control from port 2 or 4 reverses the circulation direction.

\begin{figure*}
\centering
\includegraphics[width=\textwidth]{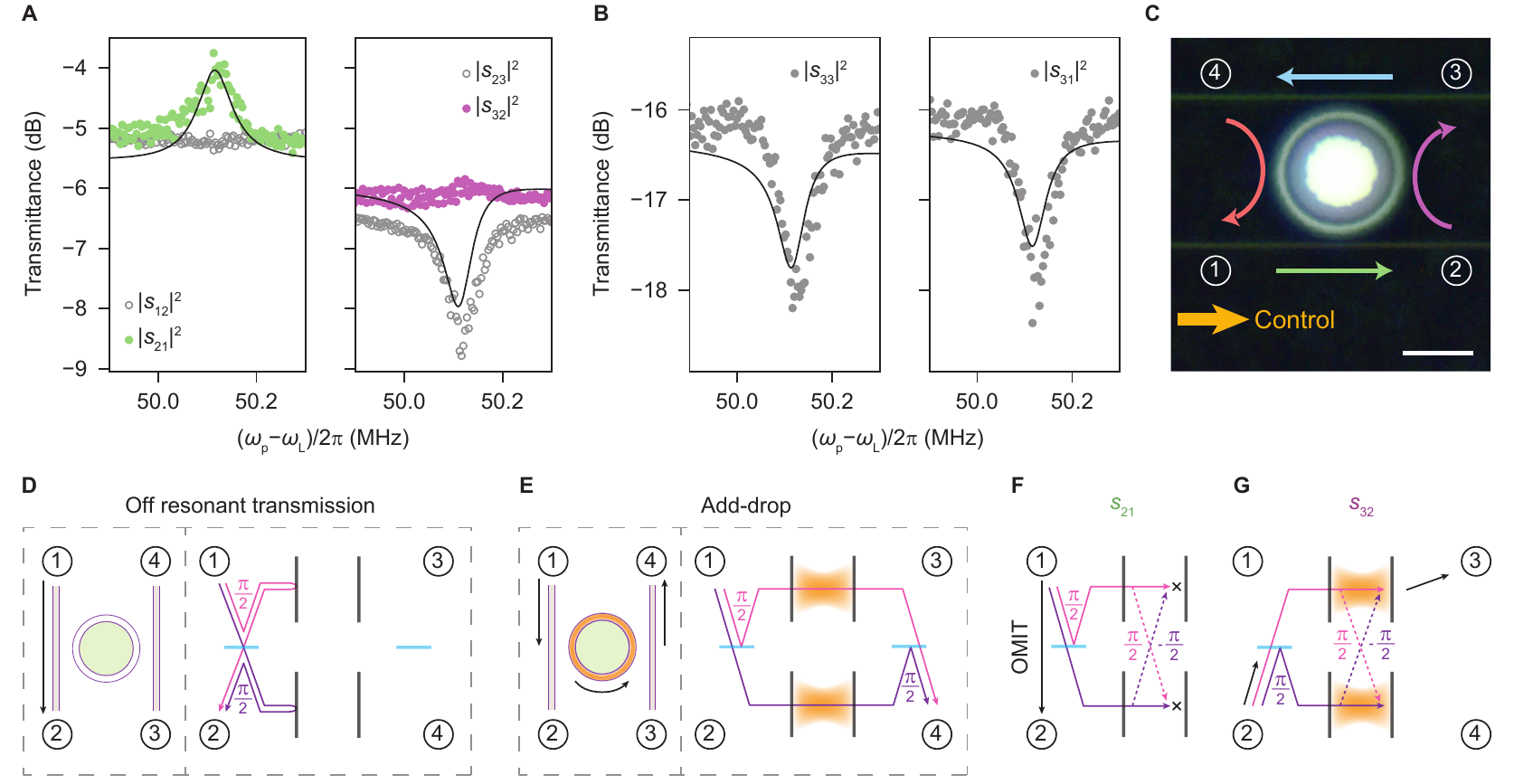}
\caption{\textbf{Nonreciprocal mode transfer and circulation.} 
(\textbf{A}) Measured probe transmittance as a function of probe-control detuning for direct (left panel) and add-drop (right panel) channels when $\Db_{1,2}\approx-\Om$. 
When probe and control beams co-propagate in the $1\leftrightarrow2$ channel (green data points), nonreciprocal optomechanically induced transparency (OMIT) is observed in both channels. 
Transmittance in directions 1$\rightarrow$2 and 2$\rightarrow$3 is larger than their reverse, yielding circulation.
(\textbf{B}) A small optical nondegeneracy yields non-zero reflection and cross-coupling, suppressed by the optomechanical interaction. 
The black lines in (\textbf{A})/(\textbf{B}) are theoretical fits for $|s_{ij}|^2$ using a global fit to all 16 transmittance spectra (Supplementary Materials). 
(\textbf{C}) Microscope image of the microtoroid and the tapered fibres (scalebar corresponds to \SI{20}{\micro\meter}), indicating ports and circulation direction. 
(\textbf{D})-(\textbf{E}) Phase pickup and corresponding constructively interfering light paths in an add-drop filter \emph{without} control beam.
(\textbf{D}) Direct transmission from $1\rightarrow2$ when the probe does not excite the optical modes.
(\textbf{E}) Add-drop functionality from $1\rightarrow4$ when the probe excites both modes with $\pi/2$ phase difference. 
(\textbf{F})-(\textbf{G}) With control beam, nonreciprocal add-drop operation is achieved using the mechanically-mediated mode transfer paths, whose opposite phases cause (\textbf{F}) destructive interference in the cavities for light entering port 1 (thus OMIT from $1\rightarrow2$), and (\textbf{G}) constructive interference in the cavities for light from port 2 (giving transmission $2\rightarrow3$).}
\label{fig:reddetuned}
\end{figure*}

In experiment, we measure circulation using a heterodyne technique (Materials and Methods) to obtain transmission spectra for all 16 port combinations that constitute the $4\times 4$ scattering matrix.
The studied microtoroid has nearly degenerate optical modes at 
$\omega_i/2\pi=\SI{196.7}{THz}$ with intrinsic loss rate %$(\kk_{0,i}=\kk_0)/2\pi=\SI{8.3}{MHz}$
$\kk_0/2\pi=\SI{8.3}{MHz}$, and supports a mechanical breathing mode at $\Omega_\mathrm{m}/2\pi=\SI{50.12}{MHz}$ ($\Gm/2\pi=\SI{62}{kHz}$).
The two optical modes are assumed to have equal total loss rates $\kk=\kk_0+\kk_\mathrm{a}+\kk_\mathrm{b}$, where $\kk_\mathrm{{a,b}}$ are exchange losses to waveguides a and b.
Figure \ref{fig:reddetuned}A shows the measured probe transmittances in the direct (1$\leftrightarrow$2) and add-drop (2$\leftrightarrow$3) channels when the control is detuned to the lower mechanical sideband of the cavity ($\Dbaronetwo\approx-\Omega_\mathrm{m}$). 
OMIT is observed in the copropagating direct channel ($|s_{21}|^2$), accompanied by reduced transmission in the add-drop channel ($|s_{23}|^2$). 
The OMIT features are absent in reverse operation of the device, which thus exhibits nonreciprocal transport. 
The overall effect is optical circulation in the direction 1$\rightarrow$2$\rightarrow$3$\rightarrow$4$\rightarrow$1 (see Supplementary Materials for all 16 scattering matrix elements).

To understand the role of mechanically-mediated mode transfer in the observed response, we recapitulate the operation of conventional add-drop filters: when a probe signal from port 1(3) does not excite the two modes (e.g. because it is off-resonance), it interferes constructively in port 2(4) and vice versa (Fig.~\ref{fig:reddetuned}D). 
When the probe does excite both modes, their $\pi/2$ phase difference means that light constructively interferes in the drop port (e.g. 1$\rightarrow$4, Fig.~\ref{fig:reddetuned}E). 
However, a control beam reconfigures this behaviour through nonreciprocal mode conversion.
Now a probe signal incident from port 1 reaches each mode either directly or via the parametric mode conversion process.
As a result of the $\pm\pi/2$ phase shift associated with the mode-transfer paths, the direct and mode-transfer path destructively interfere in both cavities (Fig.~\ref{fig:reddetuned}F). 
Consequently, a probe from port 1 does not excite the modes and is transmitted to port 2 instead of 4. 
In contrast, a probe from port 2 excites both modes with opposite phase difference, such that interference inside the cavities is constructive and light is routed to port 3 (Fig.~\ref{fig:reddetuned}G). 
Therefore, circulation can only occur in a properly configured two-mode optical system. 
In particular, the presence of a direct scattering path between ports 1 and 2 (and 3$\leftrightarrow$4) is crucial to induce the appropriate phase shift and interference causing circulation. 

The strength of the optomechanical interactions between the optical modes and the mechanical resonance is tuned by the control laser power and can be parametrised by the cooperativities $\C_i\equiv 4|g_i|^2/\kk \Gm$. 
Perfect optical circulation, with infinite isolation and zero insertion loss, is approached at high cooperativities and vanishing intrinsic loss $\kk_0$ (Fig.~\ref{fig:schematic}D). 
Stronger nonreciprocity is thus obtained for larger optomechanical coupling than currently achieved in our experiment (Fig.~\ref{fig:reddetuned}, limited by thermal instabilities), which reached 
$\C\equiv\C_1+\C_2=0.31$ (obtained from a global fit of the 4$\times$4 transmittance matrix, see Supplementary Materials). 
Moreover, the isolation could be increased by increasing the normalized coupling rates $\eta_\mathrm{a,b}=\kappa_\mathrm{a,b}/\kappa$ such that the exchange losses to the waveguides overcome the intrinsic cavity losses ($\eta_\mathrm{a,b}=(0.19,0.25)$, $\kk/2\pi=14.8$ MHz in Fig.~\ref{fig:reddetuned}).

Generally, the performance of add-drop filters is limited by a finite mode splitting $\mu=(\oo_2-\oo_1)/2$, which in ring resonators results from surface inhomogeneities that directly couple clockwise and counterclockwise propagating waves~\cite{Kippenberg2002}.
This changes the relative phase with which both (odd and even) modes are excited, leading to small ($\sim\SI{-16}{dB}$) reflection ($|s_{33}|^2$) and cross-coupling ($|s_{31}|^2$) signals in our circulator (Fig.~\ref{fig:reddetuned}B) indicative of a normalized mode splitting $\delta\equiv 2\mu/\kappa=0.19$.
Interestingly, the optomechanical interaction suppresses these unwanted signals over the nonreciprocal frequency band  (Fig.~\ref{fig:reddetuned}B).
Assuming equal control powers in the optical modes ($g_2=\im g_1$) the resonantly reflected  signal at port $j$ of waveguide a,b reads $|s_{jj}|^2=4\eta_{\mathrm{a,b}}^2 \delta^2 (1+\C+\delta^2)^{-2}$ (see Supplementary Materials).
A large cooperativity thus annihilates unwanted reflection (and likewise cross-coupling). 
This can be understood by recognizing that the nonreciprocal mode coupling at rate $\mu_\mathrm{m}$ then dominates the direct optical coupling $\mu$, making the impact of the latter on optical response negligible.

\begin{figure}
\centering
\includegraphics[width=88mm]{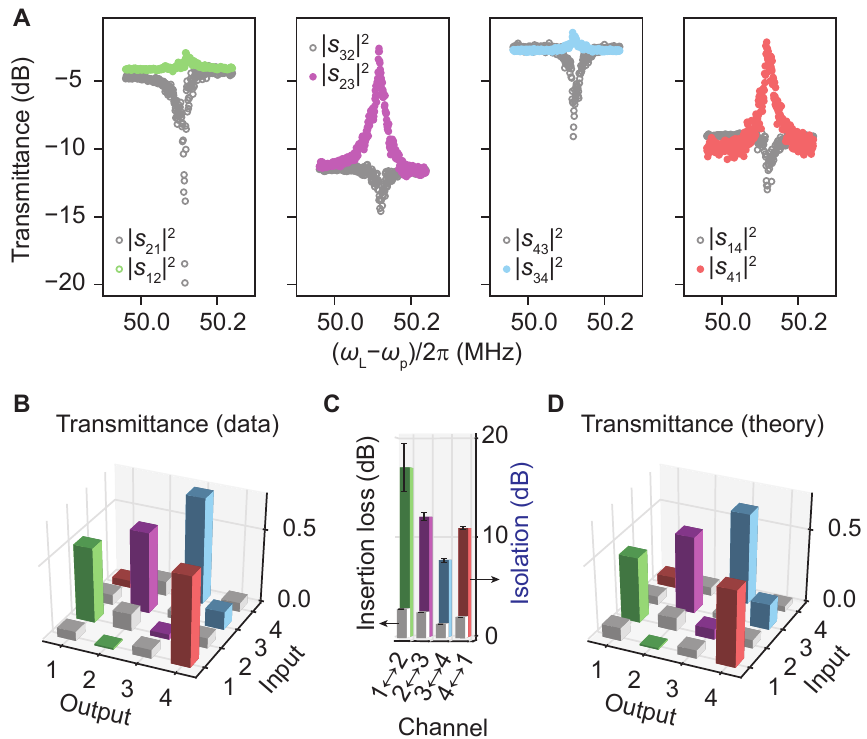}
\caption{\textbf{Circulation with optomechanical gain.} 
(\textbf{A}) Probe transmittances in the four relevant channels as a function of the control-probe detuning for $\Db=\Om$. 
The optical splitting results in  mechanically-induced absorption and transparency for probe beams co-propagating (grey data points) and counter-propagating (colored data points) to the control beam, respectively. 
Note that the circulation direction is reversed compared to Fig.~\ref{fig:reddetuned}C.
(\textbf{B}) The 4$\times$4 transmittance matrix obtained for $\octrl-\opr=\Omega_\mathrm{m}$ exhibits a clear asymmetry. 
Opposite matrix elements are the same color. 
The reflections and cross-couplings are shown in grey. 
(\textbf{C}) The corresponding isolation (colors) and insertion losses (grey).
The circulator has a maximum isolation of \SI{17}{dB} with insertion losses $<\SI{3}{dB}$ in all channels. 
Error bars represent the standard deviation. 
(\textbf{D}) Predicted 4$\times$4 transmittance matrix for a blue detuned control field when $(\kk_\mathrm{a},\kk_\mathrm{b})/2\pi=(2.8, 1.8)$ MHz, a mode splitting of $\mu/\pi=4.3$ MHz (obtained from a global fit, see Supplementary Materials) and fixed cooperativity of $\C=0.63$.}
\label{fig:bluedetuned}
\end{figure}
 
%In our proposed system, intrinsic losses limit the attainable isolation.
%However, these losses can be compensated for by introducing gain.
One way to overcome the limitation on isolation due to intrinsic loss is to introduce gain.
This naturally occurs in our system when the control beam is detuned to the blue mechanical sideband ($\Dbaronetwo\approx\Om$)~\cite{Aspelmeyer2014}.
In this regime, the device still acts as a circulator, but with opposite handedness $1\leftarrow2\leftarrow3\leftarrow4\leftarrow1$, since the mode conversion process for blue-detuned control fields has opposite phase, i.e., $\mu_\mathrm{m}=2g_1 g_2^\ast/\Gm$. 
As a result, constructive and destructive intracavity interference happen opposite to the scenarios depicted in Figs.~\ref{fig:reddetuned}F-G. 
Now, a probe signal co-propagating with the control beam experiences optomechanically induced \emph{absorption}. %, allowing for zero transmittance when $\C=1+\delta^2-2\eta_{a,b}$ (Supplementary Materials). 
Figure \ref{fig:bluedetuned}A plots the probe transmittances for the direct and add-drop channels for $\Dbaronetwo\approx\Om$.
We indeed observe a reduction of co-propagating probe signal due to induced absorption (grey data points). 
Moreover, and resulting from mode splitting, counter-propagating signals (colored data points) experience OMIT, which significantly enhances the nonreciprocal system response.
This response is characterised by inspecting the complete 4$\times$4 transmittance matrix (for $\oo_\mathrm{L}-\oo_\mathrm{p}=\Om$) shown in Fig.~\ref{fig:bluedetuned}B.
Besides displaying a clear asymmetry, we use this matrix to calculate the per-channel isolation ($\sim\SI{10}{dB}$ or more) and insertion loss ($\lesssim\SI{3}{dB}$) as displayed in Fig.~\ref{fig:bluedetuned}C.
The properties of the realized circulator agree well with the prediction of our model for a cooperativity of $\C=0.63$ (Fig.~\ref{fig:bluedetuned}D).
Importantly, our theory predicts (Supplementary Materials) that loss and gain can be balanced to give complete blocking(transmittance) in unwanted(preferred) directions.
For $\C=1$ this condition reads $\delta^2=2\eta$, resulting in a nonreciprocal bandwidth  $\Gm(1-\C/(1+\delta^2))$ and finite backreflections $\delta^2$, which can be relatively small in an undercoupled scenario.

\section*{Discussion}
\noindent{}In addition to the transport properties, the noise performance of the circulator is important for faithful signal processing, especially for applications in the quantum domain. 
Assuming equal control powers in both modes, the added noise $N_j(\omega)$ at port $j$ due to thermal occupation of the mechanical mode $\bar{n}_{\mathrm{th}}$ can be written at resonance as $N_{1,3}(\pm\Om)=4\bar{n}_{\mathrm{th}}\eta_\mathrm{a,b}\delta^2\C(1\pm \C+\delta^2)^{-2}$ and $N_{2,4}(\pm\Om)=4\bar{n}_{\mathrm{th}}\eta_\mathrm{a,b}\C(1\pm \C+\delta^2)^{-2}$, where the upper(lower) sign corresponds to a red(blue) detuned control respectively. 
Interestingly, the noise added to the ports is independent of the direction of circulation and ports 1 and 3 have noise contribution only in the presence of split optical modes $(\delta\neq 0)$. 
Active control over $\delta$ thus presents the possibility to steer noise, choosing to protect either source or receiver. 
For degenerate modes, the thermal bath adds less than 1 photon ($N_{2,4}(\pm\Om)<1$) in the optical ports when $\C(1\pm\C)^{-2}<1/(4\bar{n}_{\mathrm{th}}\eta_\mathrm{a,b})$, showing that it is beneficial to use high-frequency mechanical resonators exhibiting smaller thermal populations.
Moreover, this regime is more easily met for a red-detuned control, where an increased cooperativity widens the OMIT peak~\cite{Weis2010} and distributes the noise over a larger bandwidth.
In contrast, a blue-detuned control deteriorates this bandwidth and thus requires more stringent conditions on $\bar{n}_{\mathrm{th}}$.
%Accounting for possible mode splitting, the effective nonreciprocal bandwidth $\Gamma_\mathrm{eff}$ reads $\Gamma_\mathrm{eff}=\Gm(1\pm\C/(1+\delta^2))$ (Supplementary Materials).
%We thus observe that, for fixed cooperativity, finite splitting reduces(increases) the effective bandwidth for a red(blue)-detuned control, {\color{red}but that independent of splitting a red-detuned control gives higher bandwidths}.
As such, to achieve strong circulation at small control power, blue detuning can be beneficial, whereas to maximize bandwidth and noise performance, the red-detuned regime is preferred.
It would be good to ascertain whether schemes that include more optical or mechanical modes would allow increased flexibility to mitigate noise and improve bandwidth.  

In conclusion, we demonstrated nonreciprocal circulation of light through radiation pressure interactions in a three-mode system, with active reconfigurability regulated via the strength, phase and detuning of control fields.
Our experiments and theoretical model recognize different regimes of circulating response and reveal the general importance of destructive intracavity interference between direct and mode-conversion paths on the device properties.
These principles can lead to useful functionality for signal processing and optical routing in compact, on-chip photonic devices.

\section*{Materials and Methods}
\normalsize\noindent\textbf{Experimental details}\\
\noindent{}The microtoroid is fabricated using previously reported techniques, see e.g. \cite{Armani2003}. 
The sample is placed in a vacuum chamber operated at $3\times 10^{-6}$~mbar and room temperature. 
Two tapered optical fibres, mounted on translation stages, are positioned near the toroid to  couple the waveguides and optical modes. 
Three electronically controlled 1$\times$2 optical switches (SW) are used in tandem to launch the probe beam into one of the four ports.
A schematic of the experimental set-up is provided in the Supplementary Materials.
A tunable laser (New Focus, TLB-6728) at $\octrl$ is locked to one of the mechanical sidebands of the cavity mode, with a Pound-Drever-Hall scheme, and launched in port 1.
The output of a vector network analyser (VNA, R\&S ZNB8) at frequency $\omega$ drives a Double-Parallel Mach-Zehnder Interferometer (DPMZI, Thorlabs LN86S-FC) to generate the probe beam at frequency $\opr=\octrl\pm\omega$. 
The DPMZI is operated in a single-sideband suppressed-carrier mode, allowing selection of the relevant sideband at frequency $\opr=\octrl+\omega$ or $\opr=\octrl-\omega$. 
The power and polarisation of the control and probe beams are controlled by variable optical attenuators and fibre polarisation controllers respectively. 
The probe beam  exiting a port is combined with the control beam on a fast, low-noise photo detector whose output is analysed by the VNA. 
All four detectors are connected to the VNA through a 4$\times$1 radio frequency switch (RFSW).\\

\noindent\textbf{Scattering matrix for the optomechanical circulator}\\
\noindent{}Considering an optomechanical system that consists of two optical modes coupled to one mechanical mode, the modal amplitudes of the intracavity probe photons $\delta a_{i}$ ($i=1,2$) and the phonon annihilation operator $b$ are governed by the following coupled dynamical equations~\cite{Aspelmeyer2014}:
\begingroup
\renewcommand*{\arraystretch}{1.5}
\begin{align}
\frac{d}{dt} \delta a_{i} & =\left(\im \bar{\Delta}_{i}-\frac{\kappa_{i}}{2}\right)\delta a_{i}+\im g_{i} \left(b+b^{\dagger}\right)+\sum_{j=1}^{4}d_{ji}\delta s_{j}^{+} \label{meth1} \\
\frac{d}{dt} b & = \left(-\im\Om-\frac{\Gm}{2}\right)b+\im \sum_{i=1}^{2} g_i^*\delta a_i+g_i \delta a_i^{\dagger}+\sqrt{\Gm}b_{\text{in}}
\end{align}
\endgroup
In these relations, the loss rate $\kk_i$ of each optical mode $i$ is defined as $\kk_i\equiv\kk_{0,i} + \kk_{\mathrm{a},i} + \kk_{\mathrm{b},i}$, where $\kk_{0,i}$ refers to the intrinsic optical loss rate of the mode, and $\kk_{\mathrm{a},i}$, $\kk_{\mathrm{b},i}$ are the exchange losses to waveguides a and b. 
The $D$-matrix elements $d_{ji}$ specify the coupling between input fields $\delta s_j^+$, incident from port $j$, and the optical modes.
In addition, $g_{1,2}=g_0 \alpha_{1,2}$ and $\Db_{1,2}=\Delta_{1,2}+(2g_0^2/\Om)(|\alpha_1|^2+|\alpha_2|^2)$ respectively represent the enhanced optomechanical coupling and the modified frequency detunings, where $\Delta_{1,2}=\oo_\mathrm{L}-\oo_{1,2}$ is the frequency detuning of the control laser with respect to the optical resonance frequencies. 
Here, we consider a frequency splitting of $2\mu$ between the resonance frequencies of the two optical modes, and without loss of generality assume $\oo_{1,2}=\omega_0\mp\mu$ thus $\Db_{1,2}=\Db\pm\mu$. 
Equations (1,2) completely describe the behaviour of the system in connection with the input-output relations:
\begin{equation}\label{meth3}
%\delta s_j^{-}=\sum_{k=1}^{4}c_{jk}\delta s_{k}^{+}+\sum_{i=1}^{2}d_{ji}a_i
\begin{pmatrix}s_1^-\\s_2^-\\s_3^-\\s_4^-\end{pmatrix} = C\begin{pmatrix}s_1^+\\s_2^+\\s_3^+\\s_4^+\end{pmatrix} + D\begin{pmatrix}a_1\\a_2\end{pmatrix}
\end{equation}
where the matrix $C$ defines the port-to-port direct path scattering matrix of the optical circuit, while $D$ and its transpose $D\T$ describe the mode-to-port and port-to-mode coupling processes respectively. 
Using the rotating wave approximation, equations (\ref{meth1} - \ref{meth3}) lead to the frequency-domain scattering matrix:
\begingroup
\renewcommand*{\arraystretch}{1.5}
\begin{align}
S =& C+\im D (M+\oo I)^{-1} D\T \nonumber \\
S =& C+\im D
\begin{pmatrix}
\Soo \mp |g_1|^2/\Sm^\pm & \mp (g_1 g_2\as)/\Sm^\pm \\
\mp (g_1\as g_2)/\Sm^\pm & \Sot \mp |g_2|^2/\Sm^\pm
\end{pmatrix}^{\!\!-1}
D^\text{T}.
\label{eq:circ_generalSmatrix}
\end{align}
\endgroup
Here the upper and lower signs correspond to a control laser that is red- ($\Db=-\Om$) and blue-detuned ($\Db=+\Om$) with respect to the optical resonance. 
In addition, $\Sm^\pm\equiv\oo\mp\Om+\im\Gm/2$ and $\Sigma_{\mathrm{o}_{1,2}}\equiv\oo+\Db_{1,2}+\im\kappa_{1,2}/2$ respectively represent the inverse mechanical and optical susceptibilities, where $\Db_{1,2}=\Db\pm\mu$.
Note that $\mu_\mathrm{m}$, the mechanically-mediated mode conversion rate, is identified from the off-diagonal elements of the $M$ matrix.

The direct-path scattering matrix $C$ for the side-coupled geometry that describes both systems in Fig.~\ref{fig:schematic}A and Fig.~\ref{fig:schematic}C is
\begin{equation}
C=
\begin{pmatrix}
~ 0 ~~~~ & 1 ~~~~ & 0 ~~~~ & 0 ~ \\
~ 1 ~~~~ & 0 ~~~~ & 0 ~~~~ & 0 ~ \\ 
~ 0 ~~~~ & 0 ~~~~ & 0 ~~~~ & 1 ~ \\ 
~ 0 ~~~~ & 0 ~~~~ & 1 ~~~~ & 0 ~ 
\end{pmatrix}.
\label{eq:circ_Cmatrix}
\end{equation}
The $D$ matrix is obtained from symmetry considerations, power conservation ($D\dg D = \mathrm{diag}(\kk_\mathrm{a,1}+\kk_\mathrm{b,1},\kk_\mathrm{a,2}+\kk_\mathrm{b,2})$) and time reversal symmetry ($CD\as=-D$)~\cite{Suh2004}. 
For the even and odd mode picture, the $D$ matrix reads
\begin{equation}
D= \frac{1}{\sqrt{2}}
\begin{pmatrix}
\im \sqrt{\kk_\mathrm{a,1}} & -\sqrt{\kk_\mathrm{a,2}} \\ 
\im \sqrt{\kk_\mathrm{a,1}} & \sqrt{\kk_\mathrm{a,2}} \\ 
\im \sqrt{\kk_\mathrm{b,1}} & -\sqrt{\kk_\mathrm{b,2}} \\ 
\im \sqrt{\kk_\mathrm{b,1}} & \sqrt{\kk_\mathrm{b,2}} \\ 
\end{pmatrix}.
\label{eq:circ_Dmatrix}
\end{equation}
Equation \ref{eq:circ_generalSmatrix}, combined with eqs.~\ref{eq:circ_Cmatrix} and \ref{eq:circ_Dmatrix} fully specifies the scattering matrix of our 4-port circulator. 
We assume equal losses ($\kk_{0,1}=\kk_{0,2}=\kk_0, \kk_\mathrm{a,1}=\kk_\mathrm{a,2}=\kk_\mathrm{a}, \kk_\mathrm{b,1}=\kk_\mathrm{b,2}=\kk_\mathrm{b}$) and equal control powers ($g_2=\im g_1=\im g$) in the two optical modes to fit the experimentally measured transmittances ($|s_{ij}|^2$) using a global fitting procedure (see Supplementary Materials).

\section*{Supplementary Materials}
\noindent{}Section S1. Experimental setup\\
Section S2. Measurement and fitting procedure for the transmittance matrix\\
Section S3. Conditions for near-ideal circulation\\
Section S4. Parametric instability threshold\\
Section S5. Circulation bandwidth\\
Section S6. Thermal noise\\
Fig. S1. Experimental setup\\
Fig. S2. Data for red detuned control with fits\\
Fig. S3. Data for blue detuned control with fits\\
Fig. S4. Asymmetric transmittance matrices\\
Fig. S5. Optimal circulation\\
Fig. S6. Added noise photons\\

\noindent\\\textbf{Acknowledgments}: This work has been supported by the Office of Naval Research, with grants No. N00014-16-1-2466 and N00014-15-1-2685. It is part of the research programme of the Netherlands Organisation for Scientific Research (NWO). E.V. acknowledges support by the European Union's Horizon 2020 research and innovation programme under grant agreement No 732894 (FET Proactive HOT). \textbf{Author contributions}: F.R. developed the experimental setup. F.R. and J.P.M. performed the experiments and analysed the data. M.-A.M. developed the theoretical model, with contributions from  E.V., A.A., F.R, and J.P.M. E.V. and A.A. supervised the project. All authors contributed to the writing of the manuscript. \textbf{Competing interests}: The authors declare no competing financial interests. \textbf{Data and materials availability}: All data needed to evaluate the conclusions in the paper are present in the paper and/or the Supplementary Materials. Additional data related to this paper may be requested from the authors.

%%%%%%%%%%%%%%%%%%%%%%%
%%%%%%%%%%%%%%%%%%%%%%%%%

\renewcommand{\theequation}{S\arabic{equation}}
\renewcommand{\thesection}{Section\space{S}\arabic{section}}
\renewcommand{\thefigure}{S\arabic{figure}}

\setcounter{figure}{0}
\setcounter{equation}{0}
\clearpage
\onecolumngrid
\large
\textbf{Supplementary Materials for ``Optical circulation in a multimode optomechanical resonator"}
\normalsize

\section{Experimental setup}
\begin{figure}[h]
\centering
\includegraphics[width=88mm]{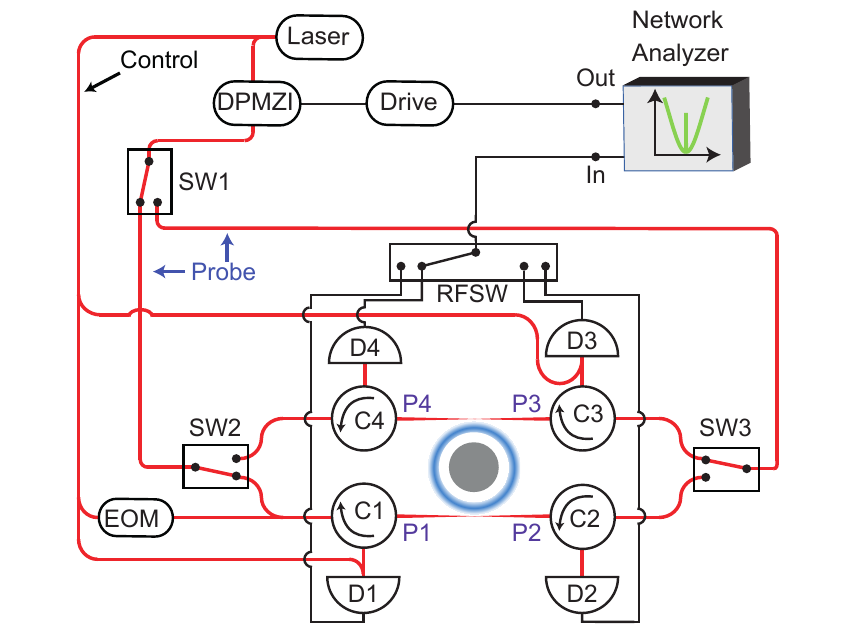}
\caption[Experimental setup]{\textbf{Experimental setup}. A detailed schematic of the experimental setup. The tapered fibres and toroid are placed in a vacuum chamber.}
\label{fig:setup}
\end{figure}

Figure~\ref{fig:setup} presents a detailed schematic of the experimental setup used in our experiment. 
Each port of the device is connected to an output of the optical switch network and to a fast detector (D) via commercially available fibre optic circulators (C, Thorlabs). 
The electronically controllable optical (SW$i$) and RF (RFSW) switches, along with the circulators, allow for straightforward measurement of the full transmittance matrix. 
Four fibre polarisation controllers (FPCs) placed directly after optical switches 2 and 3 ensure that the polarisation of the incoming probe fields can be matched to that of the cavity mode. 
The polarisation of the control field is separately controlled with a fifth FPC placed directly after the Electro-Optic Modulator (EOM).
The FPCs are omitted in the schematic for clarity.
We use the EOM for a Pound-Drever-Hall locking scheme
%~\cite{Black2001} 
that locks the control field to a motional sideband of the optical cavity. 
Furthermore, two variable optical attenuators (not shown in the schematic) placed after the DPMZI and the EOM control the power levels of the probe and control beams respectively. 
In our experiments with red- and blue-detuned control fields we used incident control powers of $\SI{60}{\micro\watt}$ and $\SI{214}{\micro\watt}$, respectively.

\label{sec:setup}

\section{Measurement and fitting procedure for the transmittance matrix}
\label{sec:fitting}
To provide calibrated and reliable $S$-matrix elements, the following experimental protocol is followed: first, careful characterization of all the lossy elements between the microtoroid and detectors is performed.
This includes all detector responses and losses associated with the commercial components and the tapered fibres. 
Next, both fibres are gently moved towards the microtoroid until the desired coupling strength is obtained. 
The latter step involves continuous spectroscopy over a frequency region around $\oo_\mathrm{c}$, and is performed at higher probe power levels ($\sim\SI{1}{\micro\watt}$) to allow direct measurements of all detector output voltages on an oscilloscope.
These voltages are obtained by splitting the detector output lines just before the RFSW (not shown).
For the direct channels ($1\leftrightarrow2, 3\leftrightarrow4$) it is then possible to use the measured off-resonance voltages to normalize the on-resonance cavity response, which directly yields the on-resonance transmittance efficiency.

\begin{figure}
\centering
\includegraphics[width=88mm]{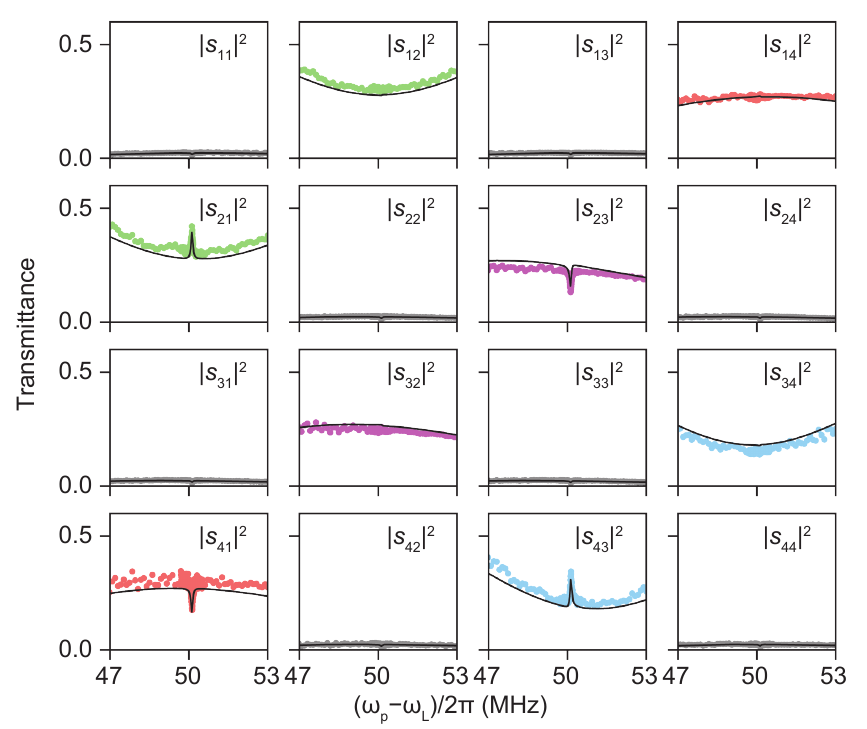}
\caption[Data for red detuned control with fits]{\textbf{Data for red detuned control with fits}. The 4$\times$4 transmittance matrix spectra that complement the red-detuned control data shown in the manuscript. The lines are global fits to the data set following the procedure described in \ref{sec:fitting}. The fitting procedure yields $(\kk_\mathrm{a}, \kk_\mathrm{b})/2\pi=(2.8,3.7)$ MHz, mode splitting of  $\mu/\pi=2.8$ MHz and cooperativity of $\C\approx0.31$.}
\label{fig:reddetuned_all}
\end{figure}

\begin{figure}
\centering
\includegraphics[width=88mm]{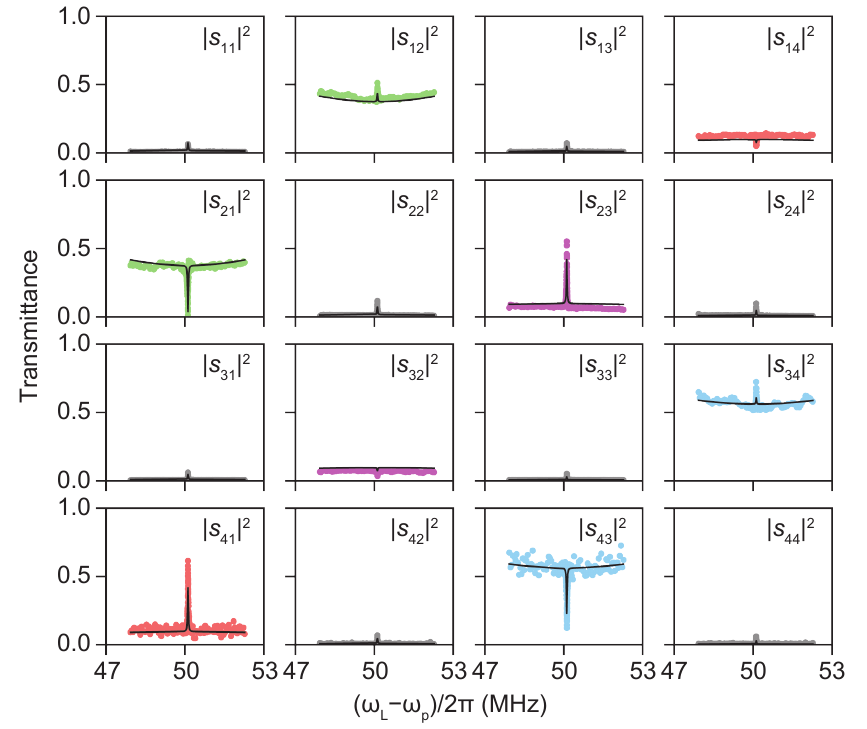}
\caption[Data for blue detuned control with fits]{\textbf{Data for blue detuned control with fits}. The 4$\times$4 transmittance matrix spectra that complement the blue-detuned control data shown in the manuscript along with theoretical fits. 
The fitting procedure yields $(\kk_\mathrm{a}, \kk_\mathrm{b})/2\pi=(2.8,1.8)$ MHz, mode splitting of  $\mu/\pi=4.3$ MHz and cooperativity of $\C\approx0.57$.}
\label{fig:bluedetuned_all}
\end{figure}

Considering for example port 1 $\rightarrow$ port 2, the non-resonant voltage on D2, together with the response of the detector and losses of circulator C2 and tapered fibre, allows us to infer the actual power that entered port 1. 
Using the on-resonance voltages that are recorded by detectors D1/D3/D4 (obtained via fitting a Lorentzian lineshape) and compensating with the appropriate losses and detector response functions, we can then 1) determine the power exiting ports 1/3/4 and 2) calculate the reflection/cross-coupling/add-drop efficiencies. 
Raw VNA spectra are converted into transmittance spectra with the help of previously determined transmittance efficiencies.  
This post-processing involves the removal of a small portion of raw VNA spectroscopic data surrounding the mechanical resonance peak.
Next, Lorentzian lineshapes are fitted to this cropped (and squared) VNA data, of which the fitted maxima are used to normalize the reflection, cross-coupling and add-drop spectra. 
For the direct channels, normalization is performed with respect to the minimum of the fitted line. 
All curves are then multiplied with their respective transmittance efficiencies to yield experimental data. 
Once the transmittances are obtained, the data around the mechanical resonance peak is included and fitted by varying the mode splitting and cooperativity as fit parameters. During fitting, $\Om,\Gm$ and $\kk_0$ are held fixed, as they are obtained from independent spectroscopic measurements. 
To account for experimental drift during data acquisition, the only parameters that are allowed to vary between the different $s$-parameters are the control detunings $\Db$.
The transmittance spectra for the full $4\times 4$ $S$ matrix, including global fit, are shown in Fig.~\ref{fig:reddetuned_all}(\ref{fig:bluedetuned_all}) for red(blue)-detuned control beam, respectively.

\begin{figure}
\centering
\includegraphics[width=88mm]{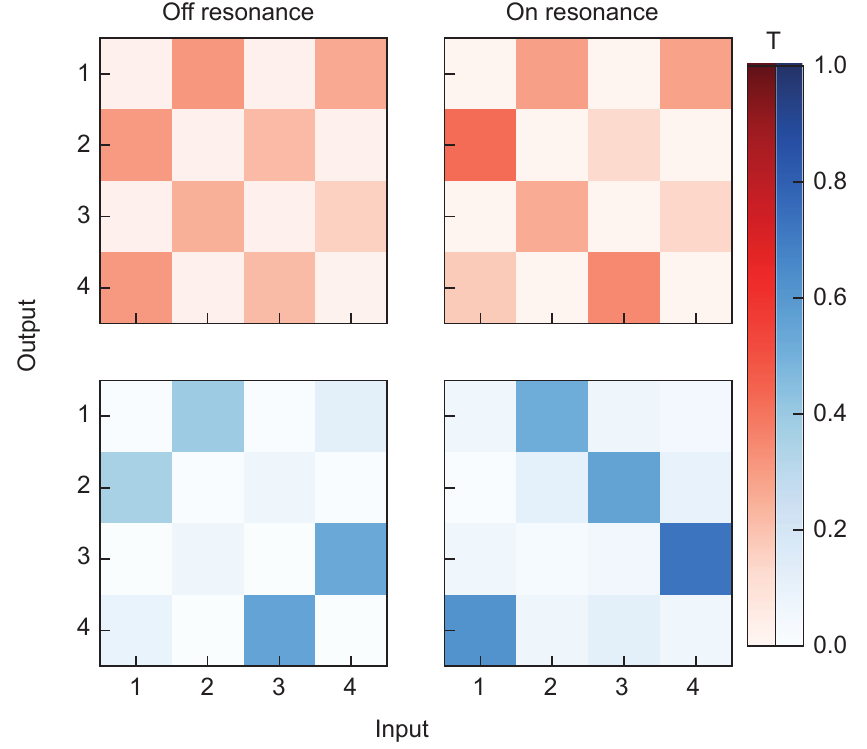}
\caption[Asymmetric transmittance matrices]{\textbf{Asymmetric transmittance matrices}. The colormaps show the transmittance values when the probe detuning frequency ($|\opr-\octrl|$) is on (right column) and off (left column) mechanical resonance. 
The top row corresponds to the case of a red detuned control while the bottom row is that of blue detuning. 
The symmetric transmittance matrix (off-resonance) becomes asymmetric when optomechanical interactions (on-resonance) are present. 
Especially for the blue detuned data the performance is close to an ideal circulator.}
\label{fig:colormaps}
\end{figure}

\section{Conditions for near-ideal circulation}
Here, we examine the scattering matrix of the circulator both in the red- and blue-detuned regimes and investigate the possibility of near-ideal circulation. For simplicity we assume that both the optical modes have equal losses ($\kk_{\mathrm{a},1}=\kk_{\mathrm{a},2}=\kk_\mathrm{a}, \kk_{\mathrm{b},1}=\kk_{\mathrm{b},2}=\kk_\mathrm{b}, \kk_{0,1}=\kk_{0,2}=\kk_0$ and $\kk=\kk_0+\kk_\mathrm{a}+\kk_\mathrm{b}$) and are pumped with equal intensity and with $\pi/2$ phase difference ($g_2=\im g_1 = \im g$). In addition, we focus only at the resonance frequency where the optomechanical intercations are maximal. For a red-detuned system ($\Db=-\Om$) at resonance ($\omega=\Om$) the $S$ matrix can be simplified to
\begin{equation} \label{Sred}
S=
\begin{pmatrix}
0&1&0&0 \\1&0&0&0 \\ 0&0&0&1 \\ 0&0&1&0 
\end{pmatrix}
-\frac{2}{1+\C+\delta^2}
\begin{pmatrix}
\im \eta_\mathrm{a} \delta &\eta_\mathrm{a} (1+\C)&\im \sqrt{\eta_\mathrm{a}\eta_\mathrm{b}}\delta&\sqrt{\eta_\mathrm{a}\eta_\mathrm{b}}(1+\C) \\\eta_\mathrm{a}&\im \eta_\mathrm{a} \delta&\sqrt{\eta_\mathrm{a}\eta_\mathrm{b}}&\im \sqrt{\eta_\mathrm{a}\eta_\mathrm{b}}\delta \\ \im \sqrt{\eta_\mathrm{a}\eta_\mathrm{b}}\delta&\sqrt{\eta_\mathrm{a}\eta_\mathrm{b}}(1+\C)&\im \eta_\mathrm{b} \delta&\eta_\mathrm{b} (1+\C) \\ \sqrt{\eta_\mathrm{a}\eta_\mathrm{b}}&\im \sqrt{\eta_\mathrm{a}\eta_\mathrm{b}}\delta&\eta_\mathrm{b}&\im \eta_\mathrm{b} \delta 
\end{pmatrix}
\end{equation}
where, $\C=\C_1+\C_2$ represents the total cooperativity of both modes. In addition, we have defined $\eta_\mathrm{a,b}=\frac{\kk_\mathrm{a,b}}{\kk}$ as the ratio of leakage losses to the total losses of each mode and $\delta=\frac{2\mu}{\kk}$ as the normalized frequency splitting of the even and odd modes.

Here, the aim is to maximize the clockwise port-to-port coupling coefficients, i.e., $\{s_{21},s_{32},s_{43},s_{14}\}$, while minimizing the counterclockwise port-to-port couplings $\{s_{12},s_{23},s_{34},s_{41}\}$. 
On the other hand, of interest would be to simultaneously minimize the reflection coefficients ($s_{11},s_{22},s_{33},s_{44}$) and the cross-coupling terms ($s_{13},s_{31},s_{24},s_{42}$). 
According to Eq. (\ref{Sred}), ideal circulation can be acheieved in the limit of large cooperativities ($\C \rightarrow \infty$) and for zero internal losses when assuming equal mode coupling to the upper and lower waveguide channels ($\eta_\mathrm{a,b} \rightarrow 1/2$). 
On the other hand, another interesting observation in the $S$ matrix of Eq.~(\ref{Sred}) is that the reflection coefficients and the cross-coupling terms are all proportional with the frequency splitting 2$\mu$ (appearing as $\delta$ in the $S$ matrix). 
This can be understood easily in the basis of rotating whispering gallery modes. In such picture, any coupling between the counter-rotating modes is mediated through their mutual coupling rate $\mu$ which instead results in a finite reflection and coupling of the diagonal ports.

%Shall we show theory plots here?% I don't think it is necessary.

\begin{figure}
\centering
\includegraphics[width=88mm]{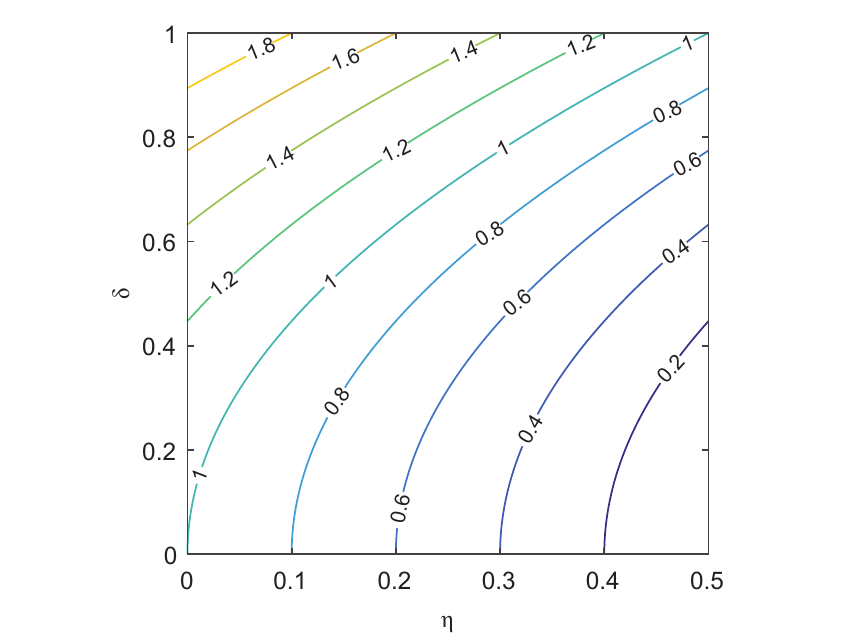}
\caption[Optimal circulation]{\textbf{Optimal circulation}. The required cooperativity for optimal circulation in the blue-detuned regime in a parameter map of $\eta$ and $\delta$.}
\label{fig:map1}
\end{figure}

Similarly, for a blue-detuned control ($\Db=\Om$) at resonance ($\omega=-\Om$) the $S$ matrix can be written as
\begin{equation} \label{Sblue}
S=
\begin{pmatrix}
0&1&0&0 \\1&0&0&0 \\ 0&0&0&1 \\ 0&0&1&0 
\end{pmatrix}
-\frac{2}{1-\C+\delta^2}
\begin{pmatrix}
\im \eta_\mathrm{a} \delta &\eta_\mathrm{a} (1-\C)&\im \sqrt{\eta_\mathrm{a}\eta_\mathrm{b}}\delta&\sqrt{\eta_\mathrm{a}\eta_\mathrm{b}}(1-\C) \\\eta_\mathrm{a}&\im \eta_\mathrm{a} \delta&\sqrt{\eta_\mathrm{a}\eta_\mathrm{b}}&\im \sqrt{\eta_\mathrm{a}\eta_\mathrm{b}} \\ \im \sqrt{\eta_\mathrm{a}\eta_\mathrm{b}}\delta&\sqrt{\eta_\mathrm{a}\eta_\mathrm{b}}(1-\C)&\im \eta_\mathrm{b} \delta&\eta_\mathrm{b} (1-\C) \\ \sqrt{\eta_\mathrm{a}\eta_\mathrm{b}}&\im \sqrt{\eta_\mathrm{a}\eta_\mathrm{b}}\delta&\eta_\mathrm{b}&\im \eta_\mathrm{b} \delta 
\end{pmatrix}.
\end{equation}
which is similar to that of the red-detuned system when replacing $\C$ with $-\C$. In a similar fashion, one can show that in this case a large cooperativity results in ideal circulation, however, this is an unphysical scenario given that in the blue-detuned regime large cooperativities give rise to parametric instabilities. On the other hand, the S matrix of Eq. (\ref{Sblue}) exhibits interesting properties and can become close to that of an ideal circulator for critical choices of the parameters involved. Interestingly, compared to the red-detuned case, here the circulation direction is reversed, i.e., $\{s_{12},s_{23},s_{34},s_{41}\}$ can be close to unity while $\{s_{21},s_{32},s_{43},s_{14}\}$ are minimum. In fact, for a particular set of parameters, one can show that $s_{21}$ and $s_{43}$ become zero for
\begin{align} \label{cond1}
&\C=1+\delta^2-2\eta_\mathrm{a}\\
&\C=1+\delta^2-2\eta_\mathrm{b}
\end{align}
which clearly requires equal leakage rate for both channels $\eta_\mathrm{a,b}=\eta$. In addition, these conditions demand a total cooperativity larger than the ratio of internal loss to total losses ($\eta_0=1-2\eta$). Figure \ref{fig:map1} depicts the cooperativity required to satisfy this condition for different values of $\eta$ and $\delta$. Interestingly, the choice of $\delta^2=2\eta$ demands a cooperativity of $\C=1$ which instead results in the following resonant S matrix:
\begin{equation} \label{Sblue1}
S=
\begin{pmatrix}
-\im\delta&1&-\im\delta&0 \\0&-\im\delta&-1&-\im\delta \\ -\im\delta&0&-\im\delta&1 \\ -1&-\im\delta&0&-\im\delta 
\end{pmatrix}
\end{equation}
According to this simple relation, near-ideal circulation can be achieved for $\C=1$ and $\delta^2=2\eta$. On the other hand, one can reduce the reflection coefficients and diagonal port couplings by decreasing $\delta$ which instead requires a decrease in $\eta$. This latter scenario thus happens for a weak waveguide-cavity coupling. 

%\section{Instability threshold}
% We can find the instability threshold in the blue-detuned regime which is missing from the discussion above. In particular, the role of the frequency splitting is very important. Is such a section required?

\section{Parametric instability threshold}
For a blue-detuned control laser, the optomechanical system can enter a parametric instability regime. For a single-mode optomechanical system, the threshold cooperativity associated with the onset of instabilities is found to be $\C_{\text{th}}=1$. Similar relation holds for the microtoroid system when assuming a degenerate pair of counter rotating modes where only one of the modes is populated by the control laser. However, assuming a finite coupling $\mu$ between the two modes, this relation is no longer valid. In this case, one would expect a higher cooperativity for bringing the system to instability threshold given that the active mode is coupled to another mode which is naturally lossy. For a blue-detuned system, the dynamical equations governing the optomechanical system in the absence of input probe signals can be written as:
\begin{equation} \label{dynamics}
\frac{d}{dt}
\begin{pmatrix}
\delta a_1 \\ b^\dagger \\ \delta a_2 
\end{pmatrix}
=\im
\begin{pmatrix}
\Db_1+\im\kappa_1/2 & g_1 & 0 \\ -g_1^* & \Om+\im\Gm/2 & -g_2^* \\ 0 & g_2 & \Db_2+\im\kappa_2/2 
\end{pmatrix}
\begin{pmatrix}
\delta a_1 \\ b^\dagger \\ \delta a_2 
\end{pmatrix}
\end{equation}
Assuming $\Db_{1,2}=\Om\pm\mu$ and $g_2=\im g_1=\im g$, we consider an ansatz of $\left(\delta a_1 ~,~ b^\dagger ~,~ \delta a_2 \right)^T=\left(\alpha_1 ~,~ \beta ~,~ \alpha_2 \right)\T e^{i\Om t}e^{st}$, which results in the following cubic equation for the exponent $s$
\begin{equation}\label{characteristic}
s^3+\left(\kappa+\frac{\Gm}{2}\right)s^2+\left(\mu^2+\frac{\kappa^2}{4}+\kappa\frac{\Gm}{2}-2|g|^2\right)s+\mu^2\frac{\Gm}{2}+\frac{\kappa^2}{4}\frac{\Gm}{2}-\kappa|g|^2=0.
\end{equation}
The stability of the system can now be investigated by finding a parameter regime for which all roots of this equation fall on the left side of the imaginary axis, i.e., $\text{Real}(s)<0$. This latter can be obtained through the Routh-Hurwitz stability criterion. According to the Routh-Hurwitz criterion, a generic cubic equation of the form $s^3+a_2s^2+a_1s+a_0$ has all roots in the negative real part plane as long as $a_0>0$, $a_2>0$ and $a_1a_2>a_0$. For Eq.(\ref{characteristic}), the stability conditions are found to be:
\begin{equation}\label{stability1}
\frac{\Gm}{2}\left(\mu^2+\frac{\kappa^2}{4}\right)>\kappa |g|^2,
\end{equation}
\begin{equation}
\left(\kappa+\frac{\Gm}{2}\right)\left(\mu^2+\frac{\kappa^2}{4}+\kappa\frac{\Gm}{2}-2|g|^2\right) > \frac{\Gm}{2}\left(\mu^2+\frac{\kappa^2}{4}\right)-\kappa |g|^2,
\end{equation}
Given that in practice $\Gm \ll \kappa$, the first condition imposes a lower bound on the control power required to bring the system to instability threshold. Therefore, by rewriting Eq. (\ref{stability1}) in terms of the total cooperativity, the condition of stability can be written as $\C<1+\delta^2$, which results into the following expression for the instability threshold
\begin{equation}
\C_{\text{th}}=1+\delta^2.
\end{equation}
This relation assures that the condition of near-ideal circulation (Eq.(\ref{cond1})) is accessible without running into instability. It is worth noting that this condition is obtained under the rotating wave approximation. A more general expression can be found when considering both mechanical sidebands.

\section{Circulation bandwidth}
According to the results presented in the main text, the bandwidth of the optomechanical circulator is dictated by the linewidth of the optomechanically induced transparency (OMIT) or absorption (OMIA) window. Considering the $S$-matrix derived in the Materials and Methods section, the bandwidth can be obtained through the frequency-dependent factor of $[\det(\omega I+M)]^{-1}$,
\begin{equation}\label{deter}
\frac{1}{\det(\omega I+M)}=\frac{1}{\Sigma_\mathrm{o}^2(\omega)-\mu^2\mp 2|g|^2\Sigma_\mathrm{o}(\omega)/\Sigma_\mathrm{m}(\omega)},
\end{equation}
where, to simplify the analysis, we have assumed equal losses in both modes, $\kappa_1=\kappa_2$, and equal intensity pumping, $|g_1|=|g_2|$. In this relation the up and down signs are associated with the red- and blue-detuned regimes respectively. 
Given that in general $\kappa \gg \Gm$, the inverse optical susceptibilities are fairly constant near the resonance $\omega=\pm\Om$ and over the small frequency range of OMIT/OMIA. 
Thus one can assume $\Sigma_\mathrm{o}(\omega)\approx \im\kappa/2$ which greatly simplifies Eq.(\ref{deter}) to
\begin{equation}
\frac{1}{\det(\omega I+M)}=-\frac{4}{\kappa^2(1+\delta^2)}\left(1-\im\frac{\pm\frac{\C}{1+\delta^2}}{\omega\mp\Om+\im\left(1\pm\frac{\C}{1+\delta^2}\right)\frac{\Gm}{2}}\right).
\end{equation}
According to this relation, the transparency or absorption window is found to be
\begin{equation}\label{bandwidth}
\text{BW}=\left(1\pm\frac{\C}{1+\delta^2}\right)\Gm,
\end{equation}
which, could be compared with a similar relation for a single-mode optomechanical cavity: $\text{BW}=(1\pm\C)\Gm$. Recalling that $\delta=\mu/(\kappa/2)$ is a normalized mode splitting, relation (\ref{bandwidth}) shows that in general the lifted degeneracy of the modes decreases the bandwidth in the red-detuned regime while it increases the bandwidth in the blue-detuned case. In addition, relation (\ref{bandwidth}) is in agreement with the fact that the bandwidth should approach zero at the onset of parametric instabilities.

\section{Thermal noise}
%Change from classical notation for consistency with main manuscript?
Thermal phonons in the mechanical resonator interact with the control beam creating cavity photons that contribute to noise at the output ports of the device. The effect of thermal noise can be considered in the linearized mechanical equation of motion as:
\begin{equation}
\frac{d}{dt}b = \left(-\im\Om -\frac{\Gm}{2}\right)b + \im \left(g_1^*\delta a_1+g_1\delta a_1^{\dagger}+g_2^*\delta a_2+g_2\delta a_2^{\dagger}\right)+\sqrt[]{\Gm}\xi(t)
\end{equation} 
where $\xi(t)$ is the noise operator with correlations $\left \langle \xi(t)\xi^{\dagger}(t') \right \rangle=(\bar{n}_\mathrm{th}+1)\delta(t-t')$ and $\left \langle \xi^{\dagger}(t)\xi(t') \right \rangle=\bar{n}_\mathrm{th}\delta(t-t')$ where $\bar{n}_\mathrm{th}=k_B T/\hbar\Om$ is the thermal occupation number of the mechanical mode. 
The coupled optomechanical equations of motion can now be used to write the frequency-domain contribution of noise to all four ports $\delta s_{\mathrm{n},i}^-(\omega)$ in terms of the noise operator $\xi(\omega)$. Under the rotating wave approximation, the output noise operators for a red-detuned system are found to be:  
\begin{equation}
\begin{pmatrix}
\delta s_{\mathrm{n},1}^-(\omega)\\ \delta s_{\mathrm{n},2}^-(\omega)\\\delta s_{\mathrm{n},3}^-(\omega)\\\delta s_{\mathrm{n},4}^-(\omega) 
\end{pmatrix}= -\frac{\im}{\Sm^+(\omega)} D(M+\omega I)^{-1}
\begin{pmatrix}
g_1\\g_2
\end{pmatrix}
\sqrt[]{\Gm}\xi(\omega),
\end{equation}
while, for a blue-detuned control laser, we have:
\begin{equation}
\begin{pmatrix}
\delta s_{\mathrm{n},1}^-(\omega)\\ \delta s_{\mathrm{n},2}^-(\omega)\\\delta s_{\mathrm{n},3}^-(\omega)\\\delta s_{\mathrm{n},4}^-(\omega) 
\end{pmatrix}= -\frac{\im}{\Sm^-(\omega)} D(M+\omega I)^{-1}
\begin{pmatrix}
g_1\\g_2
\end{pmatrix}
\sqrt[]{\Gm}\xi^{\dagger}(\omega),
\end{equation}
where,
\begin{equation}
M+\omega I=
\begin{pmatrix}
\Soo\mp\frac{|g_1|^2}{\Sm^{\pm}} & \mp\frac{g_1g_2^*}{\Sm^{\pm}}\\
\mp\frac{g_1^*g_2}{\Sm^{\pm}} & \Sot\mp\frac{|g_2|^2}{\Sm^{\pm}}
\end{pmatrix}.
\end{equation}
Here, the mechanical and optical susceptibilities are defined as $\Sm^\pm(\omega)=\omega\mp\Om+\im\Gm/2$ and $\Sigma_{\mathrm{o}_{1,2}}(\omega)=(\omega+\Dbaronetwo+\im\kk_{1,2}/2)$. 
The power spectral density of the thermal photon flux at output ports can be obtained as $N_i(\omega)=\left \langle \delta s_{\mathrm{n},i}^{\dagger}(\omega)\delta s_{\mathrm{n},i}(\omega) \right \rangle$. 
To simplify the analysis, we assume that the optical modes have equal losses ($\kk_\mathrm{a,1}=\kk_\mathrm{a,2}=\kk_\mathrm{a}$, $\kk_\mathrm{b,1}=\kk_\mathrm{b,2}=\kk_\mathrm{b}$, $\kk_{1,2}=\kk$, $\eta_\mathrm{a,b}=\kk_\mathrm{a,b}/\kk$) and are driven out of phase with equal intensities such that $g_2=\im g_1=\im g$. The noise contribution at each port then reads
\begin{equation}
\begingroup
\renewcommand*{\arraystretch}{1}
\begin{pmatrix}
N_1(\omega) \\ N_2(\omega) \\ N_3(\omega) \\ N_4(\omega)
\end{pmatrix}
= \frac{2\Gm |g|^2\bar{n}_\mathrm{th}}{|(\Sigma_o^2-\mu^2)\Sm^{\pm}\mp2|g|^2\Sigma_o|^2}
\begin{pmatrix}
\kk_\mathrm{a}  \mu^2 \\
\kk_\mathrm{a}  |\Sigma_o|^2 \\
\kk_\mathrm{b}  \mu^2 \\
\kk_\mathrm{b}  |\Sigma_o|^2
\end{pmatrix},
\endgroup
\end{equation}
where $k_B$ is the Boltzmann constant, and $T=$ \SI{300}{\kelvin} is the bath temperature.

\begin{figure}
\includegraphics[width=88mm]{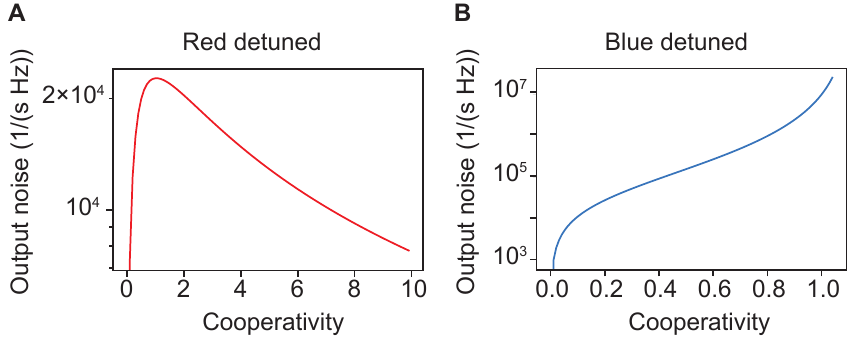}
\caption[Added noise photons]{\textbf{Added noise photons}. Maximum noise contribution to port 2 is plotted as a function of cooperativity. (\textbf{A}) For red detuned control the added noise decreases at larger cooperativities (here $\eta_\mathrm{a}=0.19$ and $\delta=0.19$). (\textbf{B}) In the case of a blue detuned control the added noise diverges when $1-\C+\delta^2=0$. Here $\eta_\mathrm{a}=0.22$ and $\delta=0.33$.}
\label{fig:noise}
\end{figure}

On resonance ($|\omega_p-\omega_L|=\Om$) the noise contribution can be written as
\begin{equation}
\begingroup
\renewcommand*{\arraystretch}{1}
\begin{pmatrix}
N_1(\pm\Om) \\ N_2(\pm\Om) \\ N_3(\pm\Om) \\ N_4(\pm\Om)
\end{pmatrix}
= 4 \bar{n}_{\mathrm{th}}\frac{\C}{(1\pm\C+\delta^2)^2}
\begin{pmatrix}
\eta_\mathrm{a} \delta^2  \\
\eta_\mathrm{a}   \\
\eta_\mathrm{b}  \delta^2 \\
\eta_\mathrm{b}  
\end{pmatrix}
\endgroup
\end{equation} where the upper and lower signs correspond to red- and blue-detuned controls respectively. Figure \ref{fig:noise} plots the output noise at port 2 for the case of red- and blue-detuned controls as a function of the cooperativity.


\begin{thebibliography}{10}

\bibitem{Sliwa2015}
K.~M. Sliwa, M.~Hatridge, A.~Narla, S.~Shankar, L.~Frunzio, R.~J. Schoelkopf,
  M.~H. Devoret, {Reconfigurable Josephson circulator/directional amplifier}.
\newblock {\it Phys. Rev. X\/} {\bf 5}, 041020 (2015).

\bibitem{Scheucher2016}
M.~Scheucher, A.~Hilico, E.~Will, J.~Volz, A.~Rauschenbeutel, {Quantum optical
  circulator controlled by a single chirally coupled atom}.
\newblock {\it Science\/} {\bf 354}, 1577--1580 (2016).

\bibitem{Lu2014}
L.~Lu, J.~D. Joannopoulos, M.~Solja{\v{c}}i{\'c}, Topological photonics.
\newblock {\it Nat. Photon.\/} {\bf 8}, 821--829 (2014).

\bibitem{Roushan2017}
P.~Roushan, C.~Neill, A.~Megrant, Y.~Chen, R.~Babbush, R.~Barends, B.~Campbell,
  Z.~Chen, B.~Chiaro, A.~Dunsworth, A.~Fowler, E.~Jeffrey, J.~Kelly, E.~Lucero,
  J.~Mutus, P.~J.~J. O'Malley, M.~Neeley, C.~Quintana, D.~Sank, A.~Vainsencher,
  J.~Wenner, T.~White, E.~Kapit, H.~Neven, J.~Martinis, {Chiral ground-state
  currents of interacting photons in a synthetic magnetic field}.
\newblock {\it Nat. Phys.\/} {\bf 13}, 146--151 (2017).

\bibitem{Shoji2014}
Y.~Shoji, T.~Mizumoto, Magneto-optical non-reciprocal devices in silicon
  photonics.
\newblock {\it Sci. Technol. Adv. Mater.\/} {\bf 15}, 014602 (2014).

\bibitem{Fang2012}
K.~Fang, Z.~Yu, S.~Fan, {Realizing effective magnetic field for photons by
  controlling the phase of dynamic modulation}.
\newblock {\it Nat. Photonics\/} {\bf 6}, 782--787 (2012).

\bibitem{Poulton2012}
C.~G. Poulton, R.~Pant, A.~Byrnes, S.~Fan, M.~J. Steel, B.~J. Eggleton, {Design
  for broadband on-chip isolator using stimulated Brillouin scattering in
  dispersion-engineered chalcogenide waveguides}.
\newblock {\it Opt. Express\/} {\bf 20}, 21235--21246 (2012).

\bibitem{Estep2014}
N.~A. Estep, D.~L. Sounas, J.~Soric, A.~Al{\`{u}}, {Magnetic-free
  non-reciprocity and isolation based on parametrically modulated
  coupled-resonator loops}.
\newblock {\it Nat. Phys.\/} {\bf 10}, 923--927 (2014).

\bibitem{Kim2015}
J.~Kim, M.~C. Kuzyk, K.~Han, H.~Wang, G.~Bahl, {Non-reciprocal Brillouin
  scattering induced transparency}.
\newblock {\it Nat. Phys.\/} {\bf 11}, 275--280 (2015).

\bibitem{Kerckhoff2015}
J.~Kerckhoff, K.~Lalumi{\`e}re, B.~J. Chapman, A.~Blais, K.~Lehnert, On-chip
  superconducting microwave circulator from synthetic rotation.
\newblock {\it Phys. Rev. Appl.\/} {\bf 4}, 034002 (2015).

\bibitem{Ranzani2015}
L.~Ranzani, J.~Aumentado, {Graph-based analysis of nonreciprocity in
  coupled-mode systems}.
\newblock {\it New J. Phys.\/} {\bf 17}, 023024 (2015).

\bibitem{Metelmann2015}
A.~Metelmann, A.~A. Clerk, Nonreciprocal photon transmission and amplification
  via reservoir engineering.
\newblock {\it Phys. Rev. X\/} {\bf 5}, 021025 (2015).

\bibitem{Aspelmeyer2014}
M.~Aspelmeyer, T.~J. Kippenberg, F.~Marquardt, {Cavity optomechanics}.
\newblock {\it Rev. Mod. Phys.\/} {\bf 86}, 1391--1452 (2014).

\bibitem{Hill2012}
J.~T. Hill, A.~H. Safavi-Naeini, J.~Chan, O.~Painter, {Coherent optical
  wavelength conversion via cavity optomechanics}.
\newblock {\it Nat. Commun.\/} {\bf 3}, 1196 (2012).

\bibitem{Andrews2014}
R.~W. Andrews, R.~W. Peterson, T.~P. Purdy, K.~Cicak, R.~W. Simmonds, C.~A.
  Regal, K.~W. Lehnert, {Bidirectional and efficient conversion between
  microwave and optical light}.
\newblock {\it Nat. Phys.\/} {\bf 10}, 321--326 (2014).

\bibitem{Hafezi2012}
M.~Hafezi, P.~Rabl, {Optomechanically induced non-reciprocity in microring
  resonators}.
\newblock {\it Opt. Express\/} {\bf 20}, 7672--7684 (2012).

\bibitem{Xu2015}
X.-W. Xu, Y.~Li, {Optical nonreciprocity and optomechanical circulator in
  three-mode optomechanical systems}.
\newblock {\it Phys. Rev. A\/} {\bf 91}, 053854 (2015).

\bibitem{Miri2017}
M.-A. Miri, F.~Ruesink, E.~Verhagen, A.~Al{\`{u}}, Optical nonreciprocity based
  on optomechanical coupling.
\newblock {\it Phys. Rev. Appl.\/} {\bf 7}, 064014 (2017).

\bibitem{Shen2016}
Z.~Shen, Y.-L. Zhang, Y.~Chen, C.-L. Zou, Y.-F. Xiao, X.-B. Zou, F.-W. Sun,
  G.-C. Guo, C.-H. Dong, {Experimental realization of optomechanically induced
  non-reciprocity}.
\newblock {\it Nat. Photonics\/} {\bf 10}, 657--661 (2016).

\bibitem{Ruesink2016}
F.~Ruesink, M.-A. Miri, A.~Al{\`{u}}, E.~Verhagen, {Nonreciprocity and
  magnetic-free isolation based on optomechanical interactions}.
\newblock {\it Nat. Commun.\/} {\bf 7}, 13662 (2016).

\bibitem{Fang2017}
K.~Fang, J.~Luo, A.~Metelmann, M.~H. Matheny, F.~Marquardt, A.~A. Clerk,
  O.~Painter, {Generalized non-reciprocity in an optomechanical circuit via
  synthetic magnetism and reservoir engineering}.
\newblock {\it Nat. Phys.\/} {\bf 13}, 465--471 (2017).

\bibitem{Peterson2017}
G.~A. Peterson, F.~Lecocq, K.~Cicak, R.~W. Simmonds, J.~Aumentado, J.~D.
  Teufel, Demonstration of efficient nonreciprocity in a microwave
  optomechanical circuit.
\newblock {\it Phys. Rev. X\/} {\bf 7}, 031001 (2017).

\bibitem{Bernier2016}
N.~R. Bernier, L.~D. T{\'{o}}th, A.~Koottandavida, M.~Ioannou, D.~Malz,
  A.~Nunnenkamp, A.~K. Feofanov, T.~J. Kippenberg, {Nonreciprocal
  reconfigurable microwave optomechanical circuit}.
\newblock {\it Arxiv 1612.08223\/}  (2016).

\bibitem{Barzanjeh2017}
S.~Barzanjeh, M.~Wulf, M.~Peruzzo, M.~Kalaee, P.~B. Dieterle, O.~Painter, J.~M.
  Fink, Mechanical on-chip microwave circulator.
\newblock {\it Arxiv 1706.00376\/}  (2017).

\bibitem{Armani2003}
D.~K. Armani, T.~J. Kippenberg, S.~M. Spillane, K.~J. Vahala, {Ultra-high-Q
  toroid microcavity on a chip}.
\newblock {\it Nature\/} {\bf 421}, 925--928 (2003).

\bibitem{Yu2009}
Z.~Yu, S.~Fan, {Complete optical isolation created by indirect interband
  photonic transitions}.
\newblock {\it Nat. Photonics\/} {\bf 3}, 91--94 (2009).

\bibitem{Fang2012a}
K.~Fang, Z.~Yu, S.~Fan, {Photonic Aharonov-Bohm effect based on dynamic
  modulation}.
\newblock {\it Phys. Rev. Lett.\/} {\bf 108}, 153901 (2012).

\bibitem{Weis2010}
S.~Weis, R.~Rivi{\`{e}}re, S.~Del{\'{e}}glise, E.~Gavartin, O.~Arcizet,
  A.~Schliesser, T.~J. Kippenberg, {Optomechanically induced transparency}.
\newblock {\it Science\/} {\bf 330}, 1520--1523 (2010).

\bibitem{Kippenberg2002}
T.~J. Kippenberg, S.~M. Spillane, K.~J. Vahala, {Modal coupling in
  traveling-wave resonators}.
\newblock {\it Opt. Lett.\/} {\bf 27}, 1669--1671 (2002).

\bibitem{Suh2004}
W.~Suh, Z.~Wang, S.~Fan, {Temporal coupled-mode theory and the presence of
  non-orthogonal modes in lossless multimode cavities}.
\newblock {\it IEEE J. Quantum Electron.\/} {\bf 40}, 1511--1518 (2004).

\end{thebibliography}
\end{document}